# The possibility of constructing a relativistic space of information states based on the theory of complexity and analogies with physical space-time

*Melnyk S.I., **Tuluzov I.G.

* O. Ya. Usikov Institute for Radiophysics and Electronics, Kharkiv, Ukraine
**Kharkiv Regional Center for Investment, Kharkiv, Ukraine

**Abstract**: The possibility of calculation of the conditional and unconditional complexity of description of information objects in the algorithmic theory of information is connected with the limitations for the set of the used languages of programming (description). The results of calculation of the conditional complexity allow introducing the fundamental information dimensions and the partial ordering in the set of information objects, and the requirement of equality of languages allows introducing the vector space. In case of optimum compression, the "prefix" contains the regular part of the information about the object, and is analogous to the classical trajectory of a material point in the physical space, and the "suffix" contains the random part of the information, the quantity of which is analogous to the physical time in the intrinsic reference system. Analysis of the mechanism of the "Einstein's clock" allows representing the result of observation of the material point as a word, written down in a binary alphabet, thus making the aforesaid analogies more clear. The kinematics of the information trajectories is described by the Lorentz's transformations, identically to its physical analog. At the same time, various languages of description are associated with various reference systems in physics. In the present paper, the information analog of the principle of least action is found and the main problems of information dynamics in the constructed space are formulated.

## Contents

## Introduction

Measurements in physics are associated with the space-time continuum, which allows solving the problems of dynamics and optimizing the operation of natural and technical systems. In this connection, numerous attempts of describing other (primarily, social-economic) systems as elements of virtual vector space and limiting the problems of prediction and optimization of their behavior to the tasks of physical dynamics have been made.

There are two main obstacles in the direction. The first is connected with the necessity of formalizing of the notions of virtual distance and intervals of virtual time between a pair of different system states. The second is connected with the presence of a subjective element in the description of their properties. This obstacle can be overcome in the theory of fundamental measurements. Previously, the model of relativistic space of economic states has been represented in our paper [1] and it has been shown that its generalization is applicable for the analysis of states of a subject's consciousness [2]. At the same time, it appeared that the presence of a subjective component inevitably results in the necessity of application of the quantum-mechanical formalism for the description of the dynamics of such systems [3, 4]. The obtained analogs between the behavior of physical and social-economic systems occur not as a result of some hidden physical mechanisms of functioning of the latter, but as a result of the general information-measurement approach to the construction of the vector space of their states [5].

In the present paper we will show that both physical and social-economic vector spaces are particular cases of the information space of the fundamental measurements and reflect all the properties and special features of the latter. By defining the fundamental measurement, we not only set the symmetries of the associated space, but also, to a significant degree, define the properties and laws of dynamics of the objects observed in this space. We assume that the consecutive development of the general theory of the information space of the fundamental measurements will result in the new results in physics, economics and in the theory of consciousness. In the present paper, we will only note some non-trivial analogs between the theory of information and physics and will show that a number of "laws of nature", which are usually considered fundamental; turn out to be the result of natural requirements imposed on the properties of the fundamental measurements.

### 1. Relative and absolute fundamental measurements in physics, in the theory of information, in economics and in the theory of consciousness

We will consider the fundamental measurement as a procedure of comparison of two objects of a certain set and the result of such measurement – their ordering relation. This definition is somewhat wider compared to the definition of the fundamental measurements as a "comparison with an etalon", proposed by N. Bohr [6]. Comparison with an etalon presupposes one of two answers: objects are "identical" or "different". Therefore the whole set of the discussed objects is divided into non-overlapping subsets of equivalent elements. Further ordering of these subsets is possible only according to the quantity of contained elements. If as a result of a fundamental measurement we obtain an ordering relation of two compared elements of the set, then the aggregate of all such results will give us a directed graph with a much c more complex structure in mathematical sense.

*In the physical theory* such property is exhibited by the relation of time ordering of any two events. As is shown in the special theory of relativity, for space-like events the result of such ordering (the answer to the question which event occurred earlier) can depend on the selection of the reference system. However, in physics, like in any other science, the aim is to extract the objective properties of the observed system invariant to the selection of the method of observation [7]. Such properties are considered objective. This allows writing down the laws of dynamics in the invariant form and then use them for a specific method of description with account of its special features. At the same time, only those connections from the set of relations of complete ordering are left, which are the same for all methods of ordering from the considered set. As a result, a partial but objective ordering of events is obtained. In physics it determines their absolute time order. Thus, for any pair of events "A" and "B" one of the following three statements is valid:

- Event «A» for any observer occurs later than event "B" (is in the upper light cone in relation to it)
- Event "A" for any observer occurs earlier than event "B" (is in the lower light cone in relation to it)
- Events "A" and "B" are space-like and can have different time order for different observers

One of these three results will be referred to as the result of the ***absolute fundamental measurement*** in the physical theory. In order to obtain it, a certain mechanism is required, which has been discussed on repeated occasions in numerous scientific works studying the essence of time. However, we will assume that the results of the absolute fundamental measurements are primary. The main mathematical idea of the method of fundamental measurements is in the absence of the necessity of their substantiation. In metaphysics this relation of the absolute time ordering is often associated with the possibility of influence of the results of one event on another (with their cause and effect relations).

The point of principal importance for our aims is the fact that the relations of partial time ordering (obtained in a particular reference system) are not measured, but calculated on the basis of the results of the absolute fundamental measurements. It becomes obvious if we accept the hypothesis on the transfer of information with the velocity of light identical in all reference systems. Then all the events about which the observer "A" has received information (already "seen" by him) are in his absolute past. But only these events are the grounds for introducing the time order of the rest of the events. Further we will show that a set of results of absolute fundamental measurements is also sufficient for introducing distances between simultaneous events, for constructing a vector space of events and for defining the reference systems as trajectories in this space.

*Referring to the theory of information,* we can note that a similar property of partial ordering is exhibited by the relation "more-less" for the quantity of information contained in information objects. However, similarly to physics, the result of such fundamental measurement can depend on the method of calculation of the quantity of information. That is why, by analogy with physics, we will define the absolute fundamental measurement in the theory of information. Then we will show that from the measurement point of view the following statements are equivalent:

- *«for any method of calculation the quantity of information contained in "A" is larger compared to the quantity of information contained in "B",* and
- *«the information object "A" contains all the information contained in the information object "B".*

Thus, the result of the absolute fundamental measurement in the theory of information is one of the three statements:

- the information object "A" contains all the information contained in the information object "B";
- the information object "B" contains all the information contained in the information object "A";
- The information "A" and the information "B" are not completely contained in each other.

At the same time we assume that in the third case the object "A" can contain part of the information about the object "B" and vice verse. We also assume that this information is relative and depends on the method of its description and interpretation by a particular subject. However, the results of the absolute fundamental measurements in the theory of information and the associated inclusion relation are absolute by definition.

*In the economic theory* the relation of partial ordering can be introduced on the basis of the procedure of transaction considered as a fundamental measurement. For any pair of economic objects "A" and "B" we can state that the observer (subject) either agrees for a transaction on exchange of "A" for "B" or refuses it. In the first case he considers that "A" is less expensive than "B", in the second case he considers that it is more expensive. The absolute fundamental measurement in economics is introduced by analogy with the previous cases. For any pair of economic objects "A" and "B" we can make one of the following three statements:

- Any of the proprietors will agree for a transaction on exchange of the object "A" possessed by him for the object "B" (object A is absolutely less expensive than object B)
- Any of the proprietors will reject a transaction on exchange of the object "A" for the object "B" (object A is absolutely more expensive than object B)
- Agreement or refusal of the transaction of the exchange "A" for "B" can depend on the proprietor's opinion.

The fundamental measurements in economics and the relativistic space of economic states constructed on the basis of their analysis have been described in detail in our papers [1, 8]. Let us also note that by analogy with the space of economic states we can also construct a relativistic space of states of a subject's consciousness. For this purpose it is sufficient to consider his choice (of a particular action) concerning the transaction offered to him by the environment. The subject either agrees for its (does not make any attempts to avoid the expected result) or rejects it (tries to change the situation), or this choice is different for different subjects.

***Thus***, we can see that the fundamental measurements in all these areas of knowledge are represented by equivalent mathematical structures. We can assume that the knowledge of the results of various fundamental measurements allows calculating the result of any other structurally more complex measurement constructed on their basis.

**2. Theory of complexity as a basis for constructing the space of information states**

*2.1. Theory of complexity as the most general mechanism of measurement of the quantity of information*

There are three generally-accepted approaches to the introduction of the notion "quantity of information". Their comparative analysis [9] shows that the first two of them (combinatorial and probabilistic) can be considered as particular cases of the algorithmic approach. Therefore we will turn our attention to the latter.

In the algorithmic approach a set of information objects represented by one-dimensional sequences ("words") of a certain finite (binary in the simplest case) alphabet is considered. For the description of this approach we will use the terminology proposed in [10].

***The conditional complexity*** $S_{\Omega_k}(B/A)$ of the information object «B» relative to object «A» is referred to the as minimal length $]\hat{p}_i[$ of program $\hat{p}_i$, which has the following properties:

- $\hat{p}_i$ is written down as a sequence of symbols (characters) of a certain language $\Omega_k$, using in the general case the alphabet different from the alphabet in which the "words" are written down;
- $\hat{p}_i$ is applicable for all "words" – elements of the set of the considered information objects;
- $\hat{p}_i$ produces the information object "B" at the output, if the information object "A" is set at the input.

Considering (and designating) the program $\hat{p}_i$ as an operator we can write down the following:

$$S_{\Omega_k}(B/A) = min\,]\hat{p}_{i\Omega_k}[ \quad if \quad B = \hat{p}_{i\Omega_k}A \quad (1)$$

As an example of such languages we can consider the Turing machine, Post machine and others.

In the fundamental theory the notion of ***unconditional complexity*** is also introduced for consideration [11]. Formally it can be defined as a conditional complexity of an information object "B" relative to a zero information object "0"; or as a minimal length of a program producing "B" at the output in case if no information is set at the input of the program, besides the start operator.

$$S_{\Omega_k}(B) = min\,]\hat{p}_{i\Omega_k}[ \quad if \quad B = \hat{p}_{i\Omega_k}0 \quad (2)$$

*2.2. Problems and limitations of the algorithmic theory of information*

*2.2.1. Non-computability of the conditional and unconditional complexity of description*

It is obvious that both conditional and unconditional complexity depend on the selection of the programming language. In this connection, the "universal language" is introduced, in which the virtual possibility of an unambiguous "translation" of programs to any other language [12] is envisaged. Then in order to write down any program in the universal language, it is sufficient to write in the source language $\Omega_k$ and set the number of this language. At the same time, the complexity of the information object "B" in the universal language will differ from the complexity of this information object in the language $\Omega_k$ for not more than

$$\delta S_B = \log_2 N_\Omega \ll S_{\Omega_k}(B), \quad (3)$$

where $N_\Omega$ is the quantity of the used programming languages. Even this adjustment shows that the refusal from limitation of the set of "simple" languages by small $N_\Omega$ makes the algorithmic definition of the quantity of information inapplicable.

In particular, ***the theorem of non-computability of complexity*** has been established. At the same time, the proof of the theorem can be represented in the style of the proof of Gödel's theorem [13]. Therefore, the obtained result is of principal nature and is associated with the logical inconsistency of the hypothesis on the existence of a universal algorithm of calculation of complexity. Nevertheless, it is obvious that for any separate programming language both conditional and unconditional complexities are computable. In order to calculate them it is sufficient to sort out all possible programs in the order of increase of their length until the required word will be produced at the output. However, in this case the question – what part of the quantity of information required for writing down the word relates to the object itself and which part relates to the language of description – remains open.

Despite this fact, the very formulation of the "theorem of non-computability" creates an illusion of existence of certain objective (not depending on the selection of the programming language) complexity, the value of which can not be accurately calculated, but can only be approximately estimated. Moreover, a number of papers have been published, in which both upper and lower threshold of the absolute value of complexity for information object are found. However, in all these works the set of languages in which the program is written explicitly or implicitly limited. We assume that the found boundaries characterize not only the word itself, but also the condition, which the programming languages are to satisfy. Further we will show that in case of absence of limitations for the selection of the programming language the value of the unconditional complexity is not only non-computable, but also loses its sense. The authors of [13] also point to this circumstance, but as an output they propose "... to hope that with the natural choice of languages this constant will be measured in thousands or even hundreds of bits." While "... if we talk about the complexities of the order of hundreds of thousands of bits (say, for the text of a novel) or millions (say, for DNA), then it's not so important which programming language we chose."

### *2.2.2. Relativity of the algorithmic complexity*

If we refuse the limitation on the quantity of possible programming languages, then the value of both conditional and unconditional complexity in case of selecting a corresponding language can be any positive number. Let us prove that for any information objects *"A"* and *"B"* written down in binary alphabet, such programming language $\Omega_k(A;B)$ exists, in which $S_{\Omega_k}(B/A) = 0$.

Let us consider the algorithmic "machine" of the following construction:

*If a binary number "A" is set at the input of the machine, then the program written down as a binary integer "n" adds n times the number $(\overline{0.k})$ to the binary number $(\overline{0.A})$, where k is the binary number of the language, and takes the fractional part from the obtained sum.*

The operation of such "machine" is described by the formula:

$$\widehat{p_{n}}_{\Omega_k} A = \{\overline{0.A} + n \cdot \overline{0.k}\} \qquad (4)$$

At the same time, identical programs written down in different programming languages correspond to different actions of the "machine". Their operation can be interpreted as the following mechanism: (Fig.1):

- Before the start of calculations the reference point in the ring is located in the upper position.
- At the moment of input of the binary word *"A"* the ring turns for angle $2\pi \cdot \overline{0.A}$ clockwise.
- In the process of execution of the program with number *"n"* it turns *n* times for the angle $2\pi \cdot \overline{0.k}$, where $k$ is the binary number of the selected programming language.
- Upon execution of the program the turning angle $\alpha(rad)$ of the ring is read and recorded as a binary word *"B"*: $\ll B \gg = \overline{0.(2\pi\alpha)}$.

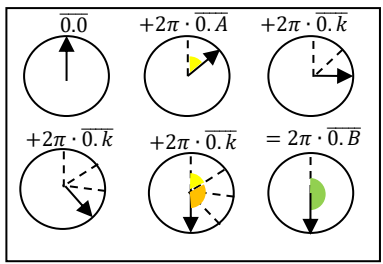

***Fig.1*** *Model of a universal "machine" for recording any binary word with the preset complexity with the selection of one of the possible languages of description.*

The mechanism of operation of the machine can be considered as a "meta language", in which a certain countable set of programming languages are written.

The length of program record equals $\log_2 n$. Then the complexity of description of the word "B" in the set language is determined by the minimal quantity of turns required for obtaining the ring position corresponding to this word.

It is easy to verify that is two binary words *"A"* and *"B"* are set, then a programming language with the number $k$: $[\overline{0.k} = \overline{0.B} - \overline{0.A} + N]$ exists, in which n=1; $N$ – integer and the relative complexity $S_{\Omega_k}(B/A) = \log_2 1 = 0$. Besides, for any set integer value $S_0$ a programming language with the number $k$: $[\overline{0.k} = (\overline{0.B} - \overline{0.A} + N)/n]$ exists, for which $S_{\Omega_k}(B/A) = [\log_2 n] + 1 = S_0$. If we use a zero word *"0"* as a word *"A"*, then for any binary word *"B"* a programming language with the number $k$: $[\overline{0.k} = \overline{0.B} + N]$ exists, in which the unconditional complexity if the word *"B"* equals 0.

Thus, the existence of a non-computable, but objective value of the absolute or relative complexity is an illusion, connected only with the a priori accepted limitations for the simplicity of the used programming languages. Let us note that the "Swift machine" from "The travels of Gulliver" also allowed obtaining with a corresponding adjustment an arbitrarily small complexity of setting an arbitrarily long text.

### *2.3. Connection of the complexity of description with the power of groups of symmetry of transformations*

Despite the above-proven statement, it seems natural that the diametrically opposite points of the ring exhibit smaller relative complexity compared to points, for instance, separated by the angle $3\pi/16$. It is due to the fact that for the language with the number ($k = 1$) the program translating A into B has the minimal length ($n = 1$). The reason of such coincidence is that the numeration of the programming languages is performed in the "natural" way. It is connected with the symmetry properties of the operator of turning for the angle $2\pi \cdot \overline{0.k}$. (The complexity of setting the number of the programming language equals to the logarithm of the power of cyclic group of rotations, the element of which is this operator).

***Thus***, we can see that the illusion of existence of a true complexity of description occurs due to those rules of numeration of different languages, which seem to us "natural" and symmetric. We will further show that at certain limitations for the properties of the used programming languages the objects and programs used for their description can be considered as elements of a vector space. Its properties are determined by the properties of those fundamental information measurements, the results of which do not depend on the selection of language (which are "objective" in the considered information space).

### 3. Information vector space

We can perform operations of addition and multiplication by natural number using a set of programs written using a certain really existing language. Thus, we will refer the program $\widehat{P}_3$ as the sum of programs $\widehat{P}_1$ and $\widehat{P}_2$, and write down $\widehat{P}_3 = \widehat{P}_1 + \widehat{P}_2$, if it is set a consecutive execution of a chain of operators of programs $\widehat{P}_1$ and $\widehat{P}_2$. Correspondingly, we will set the program $n\widehat{P}_1$ as a consecutive execution of a chain of operators of program $\widehat{P}_1$ n times. For constructing a vector space we must expand the operation of multiplication for all real (not only natural) numbers and require the execution of the following conditions (axioms):

1) *if $x, y \in V$, then $x + y = y + x$ (addition of vectors is commutative);*

2) *if $x, y, z \in V$, then $(x + y) + z = x + (y + z)$ (addition of vectors is associative);*

3) *for any $x \in V$ a vector $0 \in V$ exists (called the zero vector) satisfying the following condition $x + 0 = x$;*

4) *for any $x \in V$ a vector $y \in V$ exists (called reciprocal to x and designated as $(-x)$ satisfying the following condition $x + y = 0$;*

5) *if $x, y \in V$, and $t \in R$, then $t(x + y) = tx + ty$ (multiplication of the vector by a number is distributive relative to the addition of vectors);*

6) *if $x \in V$, and $t, s \in R$, then $(t + s)x = tx + sx$ (multiplication of the vector by a number is distributive relative to the addition of numbers);*

7) *if $x \in V$, and $t, s \in R$, then $t(sx) = (ts)x$;*

8) *if $x \in V$, then $1 \cdot x = x$.*

Let us note that for the majority of real programming languages only the axioms (2; 3; 6; 7; 8) are obviously valid. The axiom (4) on the existence of a reciprocal vector is valid if we require for any of the elementary operators of the programming language the existence of an inverse operator. For example, for the operator of writing the symbol *"1"* at the end of the binary word, the inverse operator will be the operator of deletion of the symbol "1". However, there is no unambiguously defined inverse operator for the operator of deletion of the current symbol. Thus, the introduction of the axiom (4) significantly limits the set of programming languages for which a vector space of programs can be constructed.

Axioms (1) and (5) are connected with the commutativity of programs (independence of the result from the order of their execution) and for the majority of the existing programming languages they are not valid. Besides, the operation of multiplication of programs by the real numbers additionally presupposes that any program $\widehat{P}_1$ can be represented as a sum $k$ of identical programs: $\widehat{P}_1 = k \widehat{P}_{1/k}$. In the existing programming languages the satisfaction of these properties is rather an exception than a rule.

Therefore we will further construct the set of "vector" programming languages on the basis of the theory of fundamental measurements [14], not concerning about their correspondence to the programming languages existing in reality. For this purpose we must associate the results of the fundamental measurements with the execution of programs of the considered language and their complexity. But first let us analyze the structure of an optimal program minimizing the algorithmic complexity of description of a random information object.

### 3.1. Structure of the algorithmic quantity of information and its connection with the macro- and micro-description in physical statistics

The definition of the algebraic approach to the definition of the quantity of information is based on the notion about the existence of a program of minimal length at the output of which the described information object will be obtained. In this chapter we will discuss only one fixed programming language and the quantity of information contained in the binary "word" relative to this language. As it has been shown in numerous scientific works on the theory of information any regularity contained in the recorded symbols with a sufficient length $l$ of the binary word allows limiting its description using the standard method [10].

For this purpose it is necessary to add into the program "prefix" the information on the determined regularities of arrangement of binary symbols («0»; «1») and then set in the "suffix" the binary number of one of the binary words satisfying this regularity. In the simplest case it can be the information on the quantity "1" in the word. The famous Shannon's formula can be obtained in the "algorithmic methodology" as a length of the second part of the program ("suffix"). In all cases, when the portion of "1" in the word differs from 0.5, the Shannon's formula produces the quantity of information smaller than $l$.

However, the total length of the program for writing down the word (with account of the "prefix" length) can appear to be larger than $l$. In this case we can state that the deviation of the portion "1" in the total number of symbols from 0.5 is statistically invalid and is random (fluctuation). It is easy to verify that the formula of the statistical physics for the amplitude of "equilibrium fluctuations" $\delta l \sim \sqrt{l}$ can also be obtained from the algorithmic approach to the calculation of the quantity of information. Thus, for any set information object (binary word in the considered example) we will refer the ***regularity*** as any information, which can be written in the "prefix" of the program length $l_p$, which limits the power of the set of acceptable objects by more than $2^{l_p}$ times.

With the maximal compression of a word we can find all the possible regularities in it and write them into the "prefix". Therefore, the remaining information written in the "suffix" is random by definition. This natural division of information into "regular" and "random" parts corresponds to the paradigm of description of equilibrium and non-equilibrium states in statistical physics. Thus, in the process of the description of a certain system, its macroscopic parameters are set, and then the hypothesis on equidistribution by all possible microstates of the system is used. Thus, an absolute randomness of choice of one of the possible realizations of regularities written in the macrostate is assumed, as well as the equiprobability of their occurrence in case of repeated independent choice.

### 3.2. Information structure of the results of physical measurements

The analysis of the procedure of measurements in physics has been performed in detail in many scientific works [15-17]. The most formalized scheme of measurement can be represented as a sequence of the following procedures:

- A certain device (sensor) in the set initial position is put into interaction with the observed object.
- As a result of this interaction the states of both the sensor and the object are changes in a coherent manner.
- As the interaction stops, the sensor is put into interaction with a memory cell (new in each new measurement).
- As a result of this interaction the sensor and the memory cell get into a certain equilibrium state.
- The sensor is set into its initial preset state.

In the idealized classical measurements the change of state of the observed object is neglected due to the smallness of the sensor compared to it. In quantum measurements (including continuous fuzzy measurements) this change is taken account in the full volume. For the information analysis of the procedure of changing the significant point is that as a final result the obtained (irreversibly recorded in the memory cell) information represents a set of macroscopic parameters of the final equilibrium state of the memory cell and the sensor.

The problem of information interpretation of the results of physical measurements is that most equations of physical dynamics include not these changed parameters, but the idealized functions which allow predicting the actual results (states of memory cells) with a particular accuracy. The history of development of physics, mainly in the era of domination of the deterministic outlooks, has resulted in the occurrence of an illusion that these functions are objective properties of the observed objects – those "hidden parameters", which are "in fact" possessed by physical bodies. At the same time it is considered that the results of measurements represent just a certain inaccurate description of the "hidden" objectively existing parameters.

Figure 2 illustrates the scheme of derivation of equations of physical dynamics in the classical interpretation of measurements in the framework of the information-measurement approach. In the upper part of the scheme the ***classical approach*** is illustrated. In this approach it is a priori assumed that an objective possibility of accurate description of the state of the observed object exists. The interaction of the object with the sensor results in the situation when the accurate values are imposed by additional inaccuracies. The result of measurement is formed as a result of the bifurcation process as an equilibrium irreversible state of the memory cell. Fundamental laws of physics are written down for hypothetical accurate parameters of description of the state of the observed object. They are derived as a solution of a reverse incorrect problem with account of the a priori information on the mechanism of changing and the properties of the measuring device. In these assumptions the hypothetical "accurate" parameters of the system must by independent from the properties of the used measuring device and exclude the influence of the measurement procedure.

However, the quantum mechanics has put to doubt about the possibility of such scheme of interpretation. Classical trajectory of a particle only in some idealized cases corresponds to the line of maximum possible values of the coordinate and the deviations from it can not be represented as inaccuracies of measurements. In the general case, it is distorted the stronger the more accurately this trajectory is measured. This result has been obtained in the theory of continuous fuzzy quantum measurements and it demonstrates the fact that the information obtained as a result of measurements unambiguously defines the changes of the state of the object of observation associated with it [18]. From here follows the illusion of a "freedom of choice" of a particle, which apparently "knows that it is being observed and knows what is seen" by the observers [19]. Similar information chain occurs in the theory of quantum games that shows that the behavior of classical subjects in such game with classical rules must, nevertheless, be described by the laws of quantum-mechanical formalism [20].

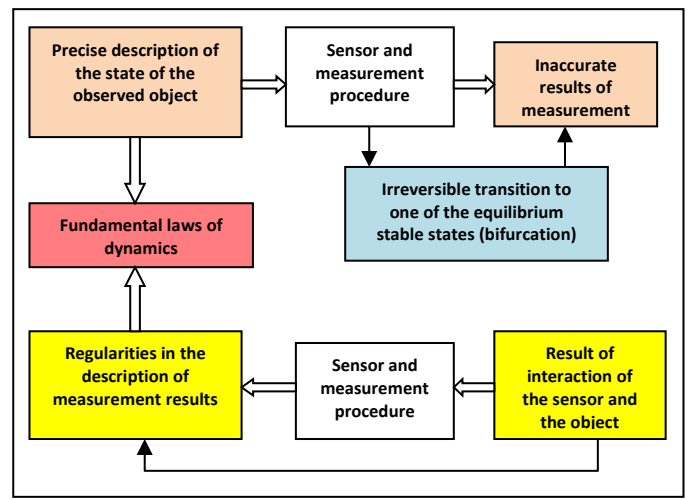

***Fig.2** Scheme of derivation of equations of physical dynamics in the classical interpretation of measurements in the framework of the information-measurement approach.*

Detailed analysis of the paradoxes and problems in modern physics which resulted in the necessity of rejecting the classical scheme has been given in the work [21]. The theory of fundamental measurements based on the ***algorithmic approach*** to the theory of information allows proposing an alternative scheme of interpretation of results of classical measurements. In the framework of this scheme, represented in the lower part of Fig.2, the result of interaction of the sensor (observer) and the observed object is initial and does not require substantiation in the form of any mechanism. It is due to the fact that any description of such mechanism requires the analysis of the results of observation of the mechanism, which require a secondary mechanism, and so on. The "circulus vituosus" occurring in this case results in numerous paradoxes caused by the attempts to include the observer into the apparatus of quantum-mechanical measurements using the mechanism of these measurements [22]. We can break this circle by considering the procedure of fundamental measurements as initial and not requiring any modeling.

The main mathematical idea of this approach is that in any information obtained as a result of measurements, regular and random parts can be sorted out. At the same time, this does not require an analysis of the mechanism of measurements (provided that it is not a part of the measurement information). The obtained separation is objective in the sense that only the random part of the information depends on the selection of the language of description (programming), while the regular part is the basis for derivation of the laws of dynamics of the observed system in an invariant form.

### 3.3. Connection of the results of fundamental measurements with the relative complexity of description

If a result of absolute fundamental measurements $A \to B$ is obtained, it means that the information "A" is contained in the information "B" at any description (unambiguously follows from it). In the theory of complexity the quantity of information $K_S(A|B)$ about «A», contained in «B», is calculated as

$$K(A|B) = S_{\Omega_k}(A) - S_{\Omega_k}(A|B), \quad (5)$$

where $S_{\Omega_k}(A)$ is the unconditional complexity of "A", which equals by definition to the quantity of information contained in "A". $S_{\Omega_k}(A|B)$ is the conditional complexity of description of "A" at the a priori set information "B". Then, however, in case of fulfillment of the condition $A \to B$ we obtain:

$$K(A|B) = S_{\Omega_k}(A), \text{ from which it follows that } S_{\Omega_k}(A|B) = 0 \quad (6)$$

The obtained result in rigorous sense contradicts to the definition of conditional complexity (1), as the complexity of programs deleting from "B" the information redundant in relation to "A", does not equal 0. For overcoming this obstacle in the theory of "complexity" the written down equalities are considered as non-rigorous, fulfilled with the accuracy up to a certain constant or a small value in relation to the summand $O(\log S(A;B))$. In this connection Kolmogorov emphasizes on repeated occasions in his works that the theory of "complexity" is applicable only for sufficiently large volumes of information. Instead of this requirement we will take account of the special features of the information-measurement approach and somewhat change the definition of the conditional complexity of description.

From the point of view of the theory of fundamental measurements we can associate the information contained in a certain object "A" with the set of answers for various questions. Thus, for instant, in order to ascertain the correctness of translation of the novel "War and peace" into Japanese language it is sufficient to compare the answers to the questions on its contents asked in two languages. At the same time, we can limit ourselves only to "fundamental" measurements – questions, for which an unambiguous answer "Yes" or "No" is possible. The more information about the novel is contained in a certain statement, the more answers to fundamental questions are unambiguously defined by this information. We will assume the complete description of a certain fundamental measurement "A" (a statement or a set of connected statements) as any other set of statements, which gives the same answers to all the questions, which can be obtained on the basis of the information "A". We will assume the conditional complexity of description of the fundamental measurement "A" in relation to "B" as the minimal quantity of results of fundamental measurements (information bit), which must be added to "B" in order to completely define "A" in the aforesaid sense.

Thus, if $A \to B$, then if "B" is set at the input of the program, we already have all the answers defining "A". Therefore, the conditional complexity of the description of "A" in relation to "B", the definition of which is adapted for the theory of fundamental measurements, equals 0. The difference from the classical definition of the conditional complexity is that we do not require the deletion of excessive information. Therefore, at the output of program $S_{\Omega_k}^*$ we obtain not exactly the object "A", but the object, which gives the same answers as "A" for the questions $F_k$ defining the object "A".

$$S_{\Omega_k}^*(B/A) = min\, ]\hat{p}_{\iota\Omega_k}\left[ if\; \forall k: F_k(B) = F_k\left(\hat{p}_{\iota\Omega_k}A\right) \right] \quad (7)$$

At the same time, the converse can appear to be false. Then the complexity of description of "B" in relation to "A" is > 0 and we consider that $A \to B$. If both relative complexities equal 0, then these are equivalent information objects (indivisible by any of the possible results of fundamental measurements).

Thus, the values of conditional complexity for a pair of objects "A" and "B" unambiguously define the relation of the absolute ordering (information inclusion) between them. However, the converse is also true. If it is known that $A \to B$, it means that the information object "B" sets unambiguous answers (Yes or No) for the same questions as "A" and additionally for some other questions. In case if there are more than one such additional answers, we can find such an object "C" that $A \to C \to B$. In the opposite case, the edge of the directed graph between "A" and "B" is an **_elementary edge_**. As the reception of an answer for one additional elementary question (fundamental measurement) requires one bit of information, we will consider the conditional complexity of such objects equal to 1. Then the conditional complexity of two random objects, for which $\to B$, can be defined as a minimal number of elementary edges in the chain $A \to C_1 \to C_2 \to C_3 \to \cdots \to B$. It is obvious that in this case the obtained value $S_{\Omega_k}^*(B/A)$ depends on the properties of the programming language $\Omega_k$.

Let us note that for any elementary edge a **_conjugate elementary edge_** of the graph must exits, corresponding to the opposite answer to the same question. In the opposite case (absence of such alternative information object and the corresponding elementary edge) the question looses its sense.

Thus, we have managed to define the results of the absolute fundamental measurements through the values of conditional complexity and vice verse: calculate the values of the conditional complexity using the results of the absolute fundamental measurements. At the same time, using the absolute fundamental measurements, we have also defined the elementary operators of the used programming language. The action of these procedures on a certain information object causes its changing for one bit of information - possibility of receiving answer for one additional question (answers "yes" and "no" correspond to different operators). Moreover, we can totally ignore the representation of the programming language operators and the information objects in the framework of the used alphabets. Instead we can define both the operators and the information objects on the basis of the elementary edges that connect them.

A similar information-measurement approach to the definition of terms is used in any sufficiently full definition dictionary. Each word in it is described using other words, which are, in turn, also described in this dictionary. Thus, a rather perspicacious reader can understand the meaning of at least some of the words by analyzing their interconnections (analog of a graph of partial ordering). At the same time, he can proceed without the a priori information on the meaning of some terms in the language he understands.

Let us note that though the definition of the absolute fundamental measurement includes the requirement of independence of their results from the selection of the language of description, they can be obtained for any of the possible languages. Thus, for instance, in physics in order to determine the result of absolute fundamental measurement between two events, it is sufficient to check if the interval between them is time-like at least in one of the reference systems.

### 3.4. Simple example of constructing the information space on the basis of a set of fundamental measurements

In classical logic any information object has an unambiguous representation in the form of a set of elementary (atomic) statements. For instance, in case of a single-time throwing of 2 coins the following elementary results are possible:

**(a) - OO; (b) – OP; (c) – (PO); (d) – (PP)**

The rest of the statements (information objects) on the results of observation can be obtained from atomic statements using logical operations. Any of them can be associated with a subset of elementary results, for which it is true. And vice verse, for any subset of elementary results a corresponding fundamental measurement exists (statement, which can possess one of two values). In the discussed example the number of such different subsets is 15 (including the set of all elements). Their structure (according to the quantity of elements of the subset) can be set as (1+4+6+4). The graph of elementary edges for these information objects is illustrated in Fig.3.

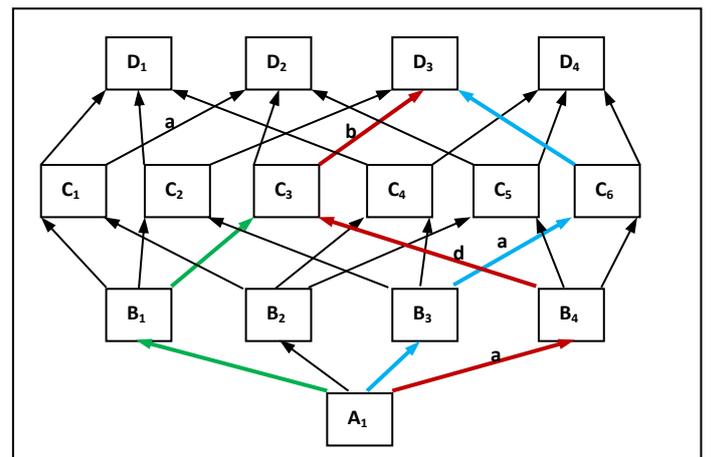

$A_1 \equiv (a \cup b \cup c \cup d)$
$B_1 \equiv (a \cup b \cup c); B_2 \equiv (a \cup b \cup d); B_3 \equiv (a \cup c \cup d); B_4 \equiv (b \cup c \cup d)$
$C_1 \equiv (a \cup b); C_2 \equiv (a \cup c); C_3 \equiv (c \cup b); C_4 \equiv (a \cup d); C_5 \equiv (b \cup d); C_6 \equiv (c \cup d)$
$D_1 \equiv (a); D_2 \equiv (b); D_3 \equiv (c); D_4 \equiv (d)$
$E_1 \equiv 0$

**Fig 3.** *Graph of elementary edges of absolute ordering of information objects and possible construction of a time scale in it.*

In this graph a sequence of elementary edges (shown in red) can be arbitrarily selected and set as a "time" scale. Relative to this scale, the rest of the graph nodes are in an unequal position. Thus, for instance, the node «$B_1$» is at the "distance" of (1+1) edges from it, while the node «$B_3$» is at the

"distance" of (1+2) edges. And if we will construct a kind of space, in which all of the node will have the corresponding coordinates, then they will be defined by this number of elementary edges connecting the mode with the "world line" taken as the time axis. We can state that the nodes «$A_1$»; «$B_4$»; «$C_3$»; «$D_3$» are at the origin of the coordinates and the rest of them are located at some distance from them. At the same time, the nodes «$C_6$» and «$C_2$» are characterized by an identical set (2+1), but contain different information. Therefore they must be located at the same distance, but in different points of the space. From this follows that even in such a simple example a one-dimensional space is insufficient for locating all graph nodes in it.

Let us note that each edge of the graph can be associated with one of the elementary events, which constitute the difference between the subsets corresponding to its ends. Some of them are designated in the illustration. Then any trajectory in the space of information spaces can be represented as a sequence of "events". For instance, the trajectory selected as a "world line" is described by the events *(a; b; d)*. At the same time, the same "event", for instance, *(a)*, can occur in different points of the information space.

We will not fix our attention on the analysis this simple example in detail, as it is very far from execution of axioms of the vector space. Let us note, however, that the introduction of an empty subset of elementary results and statements corresponding to the absence of observations makes the scheme symmetric, though more complex (1+5+10+10+5+1=32 information objects). In particular, in the expanded model for each of the elementary graph edges an opposite elementary edge appears.

The aforesaid example should be considered only as an illustration of the possibility of constructing an information space of states and introduction of a time scale and distances in it on the basis of the analysis of the results of fundamental measurements between the elements of the set of information objects. In this example we have a set of elementary results of observation a priori defined using a different language ("heads and tails") and a set of fundamental measurements constructed on its basis using logical links. The space obtained in this process is neither vector, nor contiguous, though it has a specific symmetry imposed by the initial description.

In the framework of the information-measurement approach to the construction of the space of states we, first of all, answer the question on what properties should the information objects possess so that the set of fundamental measurements would form a vector space, which would be, at the same time, maximally symmetric. Further we will discuss from these positions the properties of the procedure of ordering of events in the physical theory.

### 3.5. Space of physical events as a representation of a partial ordering of the results of absolute fundamental measurements

The classical method of constructing the relativistic space-time in physics is the analysis of light signals between observers and the use of the axiom of constancy of the velocity of light. Nevertheless, it has been shown [23] that all laws of the relativistic kinematics, including the Lorentz's transformations, can be obtained without using this postulate, and even without the analysis of light signals, but only using the requirement of equality of all inertial reference systems. The existence of a limit velocity (identical in all reference systems) of motion of material bodies is not an axiom, but a result of this analysis. Therefore, there is nothing surprising in the fact that any other space constructed on the basis of equivalent requirements of symmetry can possess the same properties.

In the physical space-time the elementary edges for the description of the absolute ordering of events are represented by linear segments of light trajectories. In the previous chapter we have shown that the quantity of elementary edges in a chain can be used as a measure of relative complexity and a prototype of a scale for the introduction of the generalized measurements. And also the latter can be introduce phenomenologically, as, for instance, in the works of Schwinger [24], we will construct their scale on the basis of the information-measurement approach.

We will consider all points of the physical continuum as a set of possible events. Then, for any two events *"A"* and *"B"* connected by a light signal such an event *"X"* can exist, that $A \to X \to B$. It means that in the classical (not quantum) physics the length of elementary edges can be made arbitrarily small and the "light trajectories" tightly fill the whole space. Moreover, any ordinary continuous trajectory can be associated with an infinite number of light trajectories, which are arbitrarily close to it. Let us note that a similar structure – "Feynman's desk" – was used as a basis for one of the interpretations of the quantum-mechanical laws of dynamics [25].

#### 3.5.1. «Einstein's clock» and its measurement-information interpretation

One of the mechanisms that allows constructing a scale of generalized measurements on the basis of the results of fundamental measurements in physics is the "Einstein's clock" - a pair of parallel mirrors, between which a beam of light is in continuous motion. On each of the mirrors or on one of them a counter of reflected light pulses must be installed. Such mechanism allows constructing a discrete time scale and determining the time of reflection of any external light pulse from the mirror surface with the accuracy up to one "tic-tac".

From the formal measurement point of view this mechanism can be interpreted in the following way:

- Two quasi-continuous sequences of absolutely ordered events «…$A$…» and «…$B$…» are set, which correspond to the trajectories of two parallel mirrors.
- On one of them an event $A_0$ is selected, which is considered as the time reference point.
- For this event another event $B_0$ is found in another sequence, such that $A_0 \to B_0$, however, no $B$ exists, such that $A_0 \to B$ and $B \to B_0$ ($B_0$ is the earliest of the events which occur absolutely later compared to $A_0$)
- For $B_0$ such an event $A_1$ is found, that $B_0 \to A_1$, however, no $A$ exists, such that $B_0 \to A$ and $A \to A_1$.
- Further, the procedure is reiterated.

Let us note that in such clock the segment of the light trajectory between the mirrors is not an elementary edge. Its length is stipulated by the randomly set distance between the mirrors and can be reduced. Nevertheless, if the distance between the mirrors becomes comparable with the length of the light wave, then the interferential effects become significant and such clock stops functioning "correctly". R. Feynman has shown [26] that for the explanation of the experimental results in this case it is necessary to refuse the classical interpretation of light trajectories in the form of a light beam. At the first stage we are going to analyze the "Einstein's clock" and their measurement-information description in the classical sense, considering that the distance between the mirrors significantly exceeds the light wave length. We will refer to the light segments corresponding to this distance as "minimal", instead of elementary, and we will consider that in all analyzed measurements the minimal graph edge is significantly larger than elementary. It assumes the possibility of consideration of a set of additional mirrors located in the middle between the main mirrors, which allow measuring the time intervals with better accuracy. In the classical approximation, such clock construction allows measuring the time intervals quasi-continuously.

In physics, there is no distinction between observers for which the distances between the mirrors in the "light clock" are different. Instead, a generalized observer is considered which can measure the moments of transmission and reception of the light signal arbitrarily accurately, i.e. this observer has a set of clocks with different (arbitrarily small) distances between the mirrors. Besides, the observer can randomly rotate them in space, measuring the turning angles. All these measurements are considered as a single generalized measurement in the observer's reference system. The problem of possibility of synchronization of a set of such clocks is not usually considered. We assume that the necessity of such *local synchronization* of the absolute fundamental measurements performed using different clocks can be used as a substantiation of the three-dimensional nature of the physical space. We will discuss this possibility in detail in our further publications.

#### 3.5.2. Distance in partially ordered sets as a method of reducing the complexity of description of the results of absolute fundamental measurements

Two stages of transition from fundamental to generalized measurements allow significantly reducing the volume of information required for the description of their results. At the first stage we left in the directed graph of partial ordering only the "elementary" or "minimal" edges. At the second stage we constructed a scale and eliminated the necessity of indicating the whole chain of intermediary events for determining the relation between the events. Now, in order to restore the result of the fundamental measurement $A \to B$, we must analyze the scale, to which both these elements belong. However, it appears that in this case each of the elements of the set can be included into the infinite quantity of various absolutely ordered chains. And only some of them are include both «$A$» and «$B$».

In order to reduce their required number to the minimum, we will use the property of transitivity. If for a certain element «$B$» an element «$A_1$» of the chain «…$A$…» is found, such that $A_1 \to B$, then this relation is true for all elements «$A_i$», such that $A_i \to A_1$. Then, it becomes obvious that a maximal element of the chain «…$A$…» exists, for which the relation is still valid. Similarly, we can determine the minimal element, for which the relation $B \to A_2$ is valid. Then, by setting two numbers: $t(A/B)_{max}$ and $t(A/B)_{min}$, we unambiguously set the results of all possible fundamental measurements between the element *"B"* and any of the elopements from the scale "A". In the physical space these two moments of time on the scale *"A"* divide it into the events appearing in the upper light cone of the event *"B"* or in the lower light cone, or appearing beyond both these cones (space-like events).

Thus, only two numbers are sufficient for setting the results of all absolute fundamental measurements between elements of an infinite sequence *"…A…"* and any other element *"B"*! However, it appears that even this information can be significantly reduced by limiting the possible constructions of the "clock" in a specific way. A similar procedure in physics is referred to as the synchronization. Using the language of the theory of fundamental measurements it means that from all possible scales we retain only those, which allow an unambiguous conversion of the values $t(A/B)_{max}$ and $t(A/B)_{min}$ into $t(C/B)_{max}$ and $t(C/B)_{min}$ for any pair of their chains «…$A$…» and «…$C$…». Moreover, the laws of conversion must be invariant to their selection. We will not quote the widely known logical conclusions and computations, which eventually result in the following definitions:

- **Distance**, calculated as $L_{AB} = \frac{1}{2}\left[t(A/B)_{min} - t(A/B)_{max}\right]$;
- **Time moment** of event «B», measured by A-clock and calculated as
  $$L_{AB} = \frac{1}{2}\left[t(A/B)_{min} - t(A/B)_{max}\right];$$
- **Inertial reference systems**, in which the clock can by synchronized;

as well as the **Lorentz's transformations**, which allow calculating the results of the generalized measurements in one reference system according to the results obtained in another reference frame and their relative velocity.

In total, they represent the most natural and consistent method of maximum simplification of the description of the set of results of absolute fundamental measurements in physics. Moreover, as it has been shown in [23], in order to derive the Lorentz's transformations in physics it is sufficient to take account pf the requirements of invariance not relying on the postulate on the constancy of the velocity of light. Therefore, there is no necessity of finding an analog of the light signals in the information or economic spaces.

Let us note that in the theory of partially ordered sets the similar notions of the maximum lower and minimum upper edge for a pair of elements "A" and "B" are considered. However, these notions are associated with the whole partially ordered set, rather than with a selected scale, unlike physical measurements.

In case of consideration of the partial ordering of a set of information objects the requirement of invariance of the description imposes a specific character on both the set of acceptable programming languages and the rules of "translation" of programs from on language into another. Further we will analyze, which parameter of information objects correspond to the parameter of physical time in the space of events.

### 3.6. Information «clock»

A scale of ordering of elements of a set, analogous to the scale of physical time, can be constructed in the theory of information as well. Formally, any sequence of information objected ordered by the relation of inclusion can be considered as such scale. Unlike the physical clock, it does not require a pair of mirrors and a light signal if a different method of calculation of the absolute and relative information contained in these objects is set. In this case, we can associate each element of the sequence with a number equal in bits to the quantity of information contained in it. Moreover, discreteness of the information scale (measurement of the "information time" in bits) is natural and does not require a separate substantiation. With such definition, there is no necessity in the synchronization of the information clock, as the value of one bit in the classical theory of information is absolute and does not depend on the selection the language of description.

In a certain sense the information scale of time is more fundamental than a scale constructed on the basis of a pair of mirrors. Let us note that in physics the mechanism of "mirror clock" is only one of the possible (but by far not the only one) methods of recording the information on the results of measurements in the form of a binary word (sequence of "0" and "1"). We can assume that in the physical theory the essence of time ordering can be observed in the mechanism of the "Einstein's clock", however it has a deeper information substantiation. In particular, let us note that the concept of the "Laplace's demon" assumes a principal possibility of possessing in any space-time point, corresponding to a certain event, the information about all events, which are lower in the light cone compared to this point. Thus, the time ordering in physics appear equivalent to the information ordering of states of the "Laplace's demon" in various points of the physical space-time.

Thus, *the information analog of time interval* between two events in the physical space separated by a time-like interval is the quantity of information required for the transition from the information object A to the information object B. In other words, it is the length of the program (in bits), in which the object A is set at the input and the object B is obtained at the output. Let us note that in physics the time interval between two events depends on the reference system in which the measurements are performed (trajectories of clock motion). In the theory of information the corresponding quantity of information depends on the programming language, which can be considered the *analog of the reference system*. Different chains of information states, each of which include all the previous ones, correspond to different trajectories of the observer's clock in physics.

Despite the aforesaid analogies, the attempts of a direct substitution of the set of physical events for a set of information objects charactering these events run into a number of principal obstacles. Further, we will analyze the reasons for which the classical theory of information does not allow forming a vector space of states (by analogy with the physical theory), and then propose its natural modification required for the occurrence of this possibility.

### 4. «Trajectory» representation of the operators of the programming language and the information objects. Algorithm of construction of the information space of a random measurement system

Let us first consider the possibility of constructing an information space on the basis of a certain real programming language. In the general case, the alphabet of such language is finite and is different from a finite alphabet used for setting the information objects (Turing machine, for instance). In order reduce it to the ***space-trajectory representation*** we can assume the following sequence of procedures:

- Determining a set of information objects with which this language is operating (for instance, a set of finite binary sequences).
- Finding for each pair of these objects a corresponding value of the conditional complexity of description (for a fixed programming language it can be done exactly).
- Constructing a set of the results of absolute fundamental measurements on the basis of the values of relative complexity for each pair of objects and single out a set of elementary directed edges from them.
- Selecting one of the nodes of the obtained directed graph as reference point and the corresponding sequence of elementary edges as a time scale.
- Determining by some or other method the distance on the basis of the results of absolute fundamental measurements and the selected scale of information time.

As a result we obtain a directed graph analogous to the graph shown in Fig.3. However, for its construction we no longer need to introduce a set of elementary events and to associate each of the information objects to a particular subset. Instead, we will use the results of the absolute fundamental measurements, which are calculated independently. A natural question arises: "Can we always represent the nodes of the graph of information states (information objects) as all possible subsets of the set of elementary events?"

In some cases the answer to this question is given by the ***Stone's theorem*** [26]. It states that for any distributive structure a monomorphism exists representing it as a structure of all subsets of a certain set in such a way, that a complement transits into a complement. A partially ordered set can be considered a structure (grid) if for any pair of its elements an exact upper and lower edge exists. If we use the results of absolute fundamental measurements as a partial ordering of the observer's states in physics, it appears that the points of intersection of the upper and lower light cones defines such edges, but only in a one-dimensional space.

In this case, the requirement of distributivity of operations of addition and multiplication is satisfied. Thus, due to the Stone's theorem, in a one-dimensional physical space the set of the observer's states can always be represented as various subsets of elementary events occurring in it. The information corresponding to the point "A" is the information state of the "Laplace's demon", which "knows" about all the events which occurred in the lower light cone of the point "A". The fact of principal importance is that for the ordering of states of the "Laplace's demon" we do not use this set of events (we are not interested in the mechanisms of his "thinking"), but do it by means of the algorithmic theory of information.

Concerning the three-dimensional physical space, the exact upper and lower edge for a random pair of events in it does not exist. Nevertheless, we assume that even in this case the proof of the generalized analog of the Stone's theorem is possible, as well as the representation of the partially ordered set of the observer's states (Laplace's demon) in the form of various subsets of elementary events ordered in the three-dimensional space-time. It is quite possible that it will require the use of quantum-mechanical formalism (in particular, the Pauli matrixes), while the elementary events will possess quantum properties (absence of localization, entangled states, etc.). A more detailed analysis of such possibility is beyond the scope of the present publication.

Returning to the analysis of the set of information states, let us note that both in the illustrative example (Fig.3) and for the existing programming languages, this set can be:

- Partially ordered on the basis of the proposed algorithm;
- Introducing an analog of the time scale;
- Introducing an analog of distance between different elements of the set;
- In some cases it is possible to represent separate edges of the graph as elementary events occurring in the information space;
- It is possible to analyze various reference systems and the connection between the descriptions of various information objects in them.

However, the obtained space is neither vector, nor continuous, nor symmetrical. Its properties are mainly determined by the properties of the used language of description. But event in this case we can note that we have a possibility of describing both the information objects and the language operators connecting them, using the time and space coordinates of the graph's nodes in the constructed information space. Thus, we obtain their ***trajectory representation***. The possibility of such representation to a significant degree makes the description of the set of information objects independent from the selection of the language of description. For instance, we can state that in case of a correct translation from English into Japanese separate sentences of any book will contain identical information. Therefore, the graphs of partial ordering constructed from these two descriptions will be identical. We can state that the meaning of messages is determined not by the selection of a particular alphabet, but by the connections of partial ordering between them and other information objects.

In order to allow the obtained results to "open" the fundamental laws of motion in the information space, let us use the "clues" from physics: B. Green writes [7]: "One of the universal lessons of the last century is that the known laws of physics are in correspondence with the principals of symmetry… For instance, why would the reference system of one observer be more preferable than another observer's system? On the contrary, from the point of view of fundamental laws of the universe it seems much more natural to interpret all reference systems equally". Considering the information space, we can also say: "Why would one language of description (programming) be more preferable than another? On the contrary, it seems much more natural to interpret all languages equally".

## 5. Information dynamics

### 5.1. Physical analog of the "prefix" and "suffix" of the compressed binary word

Let us consider the physical analogy of regular and random parts of information contained in a compressed notation of a binary word. Let us assume, for instance, that it is required to describe the classical trajectory of a material point as a result of measurements performed using the Einstein's clock. Then this description will be represented as a sequence of calculated coordinates of reflection of consecutive light signals from the surface of the observed body. Discreteness of such representation depends on the distance between the mirrors and the obtained trajectory of motion appears to be a binary "word". In this word we can single out the regular part – the frequency "1", and the random part – their location in the sequence. The regular part is defined by the relative velocity of motion of the body, while the random part is defined by the distance between the mirrors and their exact location.

Let us note that setting only the regular part of the binary word unambiguously defines the whole trajectory in any reference system. Based on this analogy, we will further consider the regular part of the compressed binary word ("prefix") as the objective part of complete information contained in this word, and the random part ("suffix") – as the subjective. At the same time the number of each symbol of the binary word corresponds to the time (number of discrete "tic-tacs") in the reference system corresponding to the selected language of description. In the "suffix" of the maximally compressed word $w(1) = w(0) = 1/2$. It means that the symbols of such language of description correspond to the intrinsic reference system, in which the body is static. In any other system $w(1) \neq w(0)$, and the length of notation of the "suffix" increases according to the Shannon's formula. It is completer analogous to the effect of relativistic dilation of physical time in moving reference systems. It can be shown that the formulas of transformation of the "prefix" and the length of the "suffix" for the set of invariant binary languages are equivalent to Lorentz's transformations in the special theory of relativity. There is nothing surprising in it, as these formulas are derived using the same principles of invariance. Thus, the rest of the laws of kinematics of the relativistic (invariant) space of binary languages of description can also be obtained. However, the solution of problems of information dynamics requires using additional principles.

### 5.2. Information analog of the principle of least action

At the end of the previous chapter we have defined the analog of the time interval between two information objects as a length of program, which converts one of them into the other. If we assume that the program is written down as a sequence of elementary operators, each of which transforms the initial information object into a certain intermediate state, then the program has a corresponding trajectory in the vector information space. In the theory of complexity from all possible programs (information trajectories) converting «A» into «B» and written in a certain fixed language, the program with minimum length is selected. This length determines the relative complexity of these objects. In physics according to the principle of least action from all possible trajectories the trajectory of motion of a free material point between two set events is selected. An obvious analogy comes to mind, which allows assuming that the analog of action (along the set trajectory) between a pair of events in the physical space is the program connecting a pair of objects in the information space. This analogy turns out to be even more obvious if we note that in physic the action between two events in the process of the motion of the material point is proportional to the time interval between them measured using the clock of a corresponding (intrinsic) reference system. However, the intrinsic time for this trajectory is maximal of all possible, while the program length for the selected information trajectory is minimal.

This apparent contradiction is connected with the fact that in the relativistic mechanics the action can be written down as:

$$S = -mc \int_a^b ds = -mc \int_a^b \sqrt{1 - \frac{v^2}{c^2}} dt = \int_a^b L dt \qquad (8)$$

$$\text{где } L = -mc^2 \sqrt{1 - \frac{v^2}{c^2}} \cong -mc^2 + \frac{mv^2}{2}$$

Thus, the action for the optimum trajectory is minimal and, at the same time, always negative and, therefore, its absolute value (the value corresponding to the intrinsic time) tends to maximum. But then the complexity of the program as an information trajectory must tend to maximum instead of minimum.

Let us note that the function of action in physics is defined by macroscopic parameters of motion (dependence of the velocity on the time for a material point). The analog of this dependence for information objects is the frequency dependence of the "letters" of the binary word, which is written into its "prefix" in case of its compression. At the same time, the value of action proportional to the intrinsic time corresponds to the length of the random part of the compressed "word". Therefore, the principle of least action corresponds in the information space not to the minimization of the complexity of description of an information object, but to the maximization of the length of the "suffix" in the maximally compressed description. This, in turn, is equivalent to the requirement of maximum quantity of random information in the program converting «A» into «B».

At the first glance, this result contradicts the main paradigm of the algebraic theory of information – the principle of minimization of complexity of description (search of a program of minimal length which converts the information object «A» into «B»). However, in case of a more detailed analysis we can solve this contradiction.

The point is that both in physical and in information space we are dealing with two different problems of optimization instead of one. In the first problem (selection of the trajectory of motion) we maximize the intrinsic time for the true trajectory. In the second – we find its simplest description transferring into the intrinsic reference system. In this case, the time calculated in the intrinsic reference system turns out to be minimal compared to the rest of the systems.

***The first variation problem*** (problem of classical dynamics) is to maximize the length of the "suffix" in the optimum trajectory record. It means that such a trajectory (program of transition from «A» to «B») is to be found, in which the minimal quantity of regularities is present. Actually, these regularities contain only the a priori information on the finite points of the trajectory and its continuity. This requirement can be associated with the principle of the "Occam's razor". In the found trajectories there are no "excess" economic entities, not following from the problem statement. This result also corresponds in the generalized sense to the principle of maximum likelihood. In our paper [28] we have shown that it can be obtained by means of the "theory of complexity". The obtained trajectory allows maximal quantity of random realizations and therefore can be considered maximally plausible.

In the ***second variation problem*** we no longer change the obtained trajectory; we are to find the simplest method of description of each of its random realizations. For this purpose, all its regularities set by the dependence $[w(1)](i)$, are transferred into the prefix, while the length of the suffix turns out to be minimal in a set of various languages of description. It is essential that the length of the random part of the optimal trajectory (information analog of action) can also be calculated for any other language of description. However, for a different language it will no longer be equal to the time interval between the two events.

### 5.3. Information analog of the interval between two information objects

Similarly to the physical space, we can introduce the notion of information interval between two information objects.

Its value for any pair of objects can be calculated in accordance with the formulas known from physics. However, it is more convenient to associate the value of the interval directly with the results of absolute fundamental measurements of these two objects in a certain reference system.

$$(\delta s_{AB})^2 = \delta T'_{AB} \delta T''_{AB}, \qquad (9)$$

where $\delta T'_{AB}$ and $\delta T''_{AB}$ are the intervals of measured time between the minimal upper and maximal lower edges of objects «A» and «B» in the set of readings of a random scale, respectively. In physics it is the time interval between the sending of two light signals and the return of their reflections from objects located in the space-time points «A» and «B». Fig.4 illustrates all three values contained in this formula.

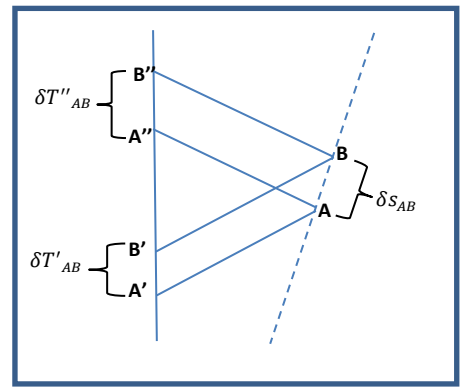

***Fig. 4*** *Determination of time interval and velocity on the basis of the results of absolute fundamental measurements*

On the other hand, the relative velocity of the observer's system and the system, in which «A» and «B» are in the same point, is calculated on the basis of the same values as

$$(v/c)^2 = \left[\frac{\delta T''_{AB}}{\delta T'_{AB}} - 1\right] / \left[\frac{\delta T''_{AB}}{\delta T'_{AB}} + 1\right] \qquad (10)$$

It is worth noting that in this case the interval $\delta s_{AB}$ is equal to the length of the «suffix», and the relative velocity $v/c$ sets the regularity which allows shortening the record of the trajectory «AB». In various languages of description the length of the random part will be identical and equal to the intrinsic information time for objects «A» and «B». The regular part can be transformed from one language into another using the Lorentz's formulas.

Let us consider in greater detail the information aspects of these two formulas. In accordance with the aforesaid, we can state that any sequence of the results of fundamental measurements (answers "Yes" or "No" for specific questions) can be interpreted as a random scale for measurement of the quantity of information. Readings of this scale contain certain information and differ by 1 bit from adjacent readings (in the assumption that this scale is accepted as the fixed reference system in the information vector space). As any scale is unlimited, we can assume that for any information object «A» sooner or later such an object «A"» will be found on the scale, which will contain all the information about «A». In the language of classical logics (logical implication) it means that «A"» is the closest reason of «A» in the selected scale or «A» is always true when «A"» is true. Correspondingly, there will be an object «A'» which will be the closest consequence of «A». Then the interval between the objects «A» and «B» in any scale represents the product of the difference in the quantity of information between their closest reasons and closest consequences. For a set of languages, which are "inertial" relative to each other (at least, in the considered interval), this product is invariant.

Thus, the value of this invariant can be considered as a certain "absolute" quantity of information separating the objects «A» and «B» and the measured values $\delta T'_{AB}$ and $\delta T''_{AB}$ – as its representations in the selected language of description.

Summing up this consideration, we can state the following:

- The problem of information dynamics is reduced to the determination of the most plausible trajectory of transition from one information object to another.
- At maximal compression (according to Vietinghoff, for instance) the description of these states consists of the "prefix" and the "suffix".
- By the trajectory of transition is meant the regular part of the program (its "prefix") written in one of the equal binary languages.
- The Lorentz's transformations in the information space allow transition from one language to another in the recording of the "prefix".
- The highest plausibility of the information trajectory corresponds to the maximal number of its random realizations or the maximal "suffix" length.
- This length is invariant to the selection of the language of description and is analogous to the value of the interval between time-like events in the physical space.
- The most plausible information trajectory is determined by solving the variation problem of maximization of the "suffix" length at the set initial and finite information states.
- This problem is equivalent to the principle of least action in the classical relativistic mechanics.
- The selection of language, in which the algorithmic complexity of the found trajectory is minimal, corresponds to the transition into the intrinsic reference system in physics.

### 5.4. Problems of information dynamics

The analogy with the principle of least action allows us proceeding to the solution of various problems of information dynamics by analogy with physics. The most important of them is the problem of prediction of the "information future" of the observed enclosed information system.

In physics this problem is formulates as the determination of the trajectory of motion of a material point representing the system in is phase space on the basis of its generalized coordinates and impulses at the initial moment of time.

#### 5.4.1. Information analog of the material point

In mechanics the material point is referred to as the system of elementary (indivisible by means of mechanics) objects, the distance between which in the considered problem can be neglected. As the information analog of the elementary objects we will consider the indivisible (by means of the selected language of description) information objects. Of course, the question on the possibility of dividing these objects into parts, i.e. representing it as two or more objects (independent or connected by information) can be solved only on the basis of analysis of the results of fundamental measurements.

If we assume that such representation is possible, then the distance between the two parts of the object will be defined by the difference in the receipt of the signals reflected from these parts. In this case we obtain the diagram shown in Fig.5. As the value $(\delta s_{AB})^2 = \delta T'_{AB} \delta T''_{AB}$ in this case is negative, the interval turns out to be imaginary, and $|\delta s_{AB}|$ characterizes the maximal distance between the points «A» and «B», simultaneous in the intrinsic reference system. Therefore, we will consider the information analog of the material point as such a system of information objects, all parts of which have indistinguishable images in a random information scale. It also means that the regular parts of information describing «A» and «B» are indistinguishably close. We can state that the corresponding "prefixes" (frequency dependences of binary words) are identical. At the same time, the random parts ("suffixes") are in no way interconnected, though they have the same length and frequency of symbols (equal to 0.5).

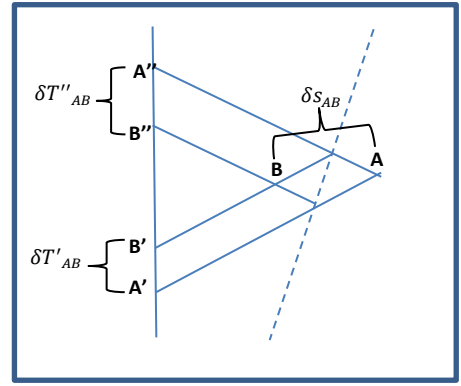

**Fig. 5** *Determination of the absolute information distance between two space-like information objects on the basis of the results of fundamental measurements.*

#### 5.4.2. Information analog of the mass of the material point

In physics the mass of the material point is represented as a coefficient of proportionality in the function of action (8). For information objects the description of the material point (as it has been shown above) can be represented as a "prefix", identical for all o its parts, describing the classical trajectory, and a "suffix" consisting of a set of absolutely random binary sequences of the same length. Then we can accept their quantity as the information mass, and the mass of one such sequence – as the unit mass.

Nevertheless, in case of such definition of the information mass, one more problem remains connected with the definition of the unit of information length (or unit of information time). In physics an etalon is used for this purpose, and only in the theories of super combination the possibility of independent calculation of the Planck's unit of length using other constants is considered. We assume that in the information space the role of this constant is performed by 1 bit of information. However, as long we consider the macroscopic information objects, we don't have grounds to consider this parameter measurable. Instead of it, the unit of length in the record of the information trajectory is set by the analog of the "Einstein's clock" and can be selected arbitrarily small (but still far from the Planck's limit).

Therefore, speaking about the mass of a macroscopic information object, we can estimate not the quantity of fundamental (Planck's) trajectories described by the same "prefix", but the relation of the quantity of information contained in the "suffix" to the quantity of information contained in the "prefix" of a compressed word. We can accept the fixed value of such ratio as the unit of mass.

#### 5.4.3. Arrow of the information time

For simplicity, we will further analyze the information analog of the information material point «A» of unit mass moving in a one-dimensional information space. Then its initial position in the phase space is defined by the information on images (moments «A'» and «A''» relative to the selected scale) and the frequency of binary symbols in a random sequence describing its trajectory (analog of information velocity).

At the same time, the fact of principal importance is that the "previous" positions of the information analog of the material point «A» are not set, but calculated on the basis of the variation dynamic principle. If we remember the famous Zeno's paradox on the "immobility of a flying arrow", we can see that in the information space it finds its natural solution. It is connected with the fact that the positions of the information analog of the material point in previous moments of information time are already contained in its current state, while the coordinates only reflect its projections on the selected scale.

Similarly, the moments of information "future" are calculated on the basis of the same information state in the "present". Returning to the analogy with the "flying arrow", we can state that the Zeno's information state at the moment of observation contains not only the coordinates of its present position, but also some additional information, which can be represented as a program satisfying the principle of maximum plausibility.

In other words, the moments of the information past of the state «A» represent the following sequence of states

$$\ldots «A_{-3}» \rightarrow «A_{-2}» \rightarrow «A_{-1}» \rightarrow «A», \qquad (11)$$

where each previous state is contained in the subsequent (in information sense) and ensures the maximal quantity of random realizations for the transition between them. That is why we can state that the "information past" is the most plausible explanation of the "information present".

Contrastingly, the "information future" is a sequence of state, where

$$\ldots «A» \rightarrow «A_1» \rightarrow «A_2» \rightarrow «A_3» \ldots, \qquad (12)$$

where each previous state is also contained in the subsequent one. At the same time, the "information present" turns out to be the most plausible explanation of the "information future". It can be shown that though only one plausible explanation of the set state "A" in the "information past" exists, this state itself can be the most plausible explanation for various states in the "information future". The asymmetry of the "past" and the "future" occurs because we construct them on the basis of the present by various methods. In order to calculate the "past" we need to erase the "excess" information from the state "A" and for calculating the "future" we need to add it.

Nevertheless, in the macroscopic limit the trajectory segments, considered as minimal (with constant velocity), contain such a large quantity of bits of information that the spreading of the information future for an enclosed system (in the absence of external forces) can be neglected. Therefore, the macroscopic dynamics of enclosed information systems, similarly to physics, turns out to be reversible. We have previously shown that at the expense of refusal of idealization of the infinitely accurate measurement the paradox of irreversibility can be analogously solved in the physical dynamics as well [29].

## 6. How to use it

The obtained analogies with physics allow expecting that in the information theory solution of the same problems as in physics would also be possible. However, the specific nature of information problems significantly differs from the problems of physical dynamics. Therefore, we should primarily reformulate the information problem statement in such a way, that it could be solved analogously to a physical problem. Let us note that the results of its solution have sense only in case when they can be verified by observation. Actually, they are nothing else but predictions of the results of fundamental measurements or their properties. These predictions are made on the basis of all available a priori information. And this a priori information can be obtained as the regular part of the set of result of already executed fundamental measurements. Further we will analyze the algorithm of application of the methods of physical dynamics for the prediction in comparison with the classical methods.

*In classical logics* the results of fundamental measurements are the values of validity of various statements. At the same time it is assumed that any of the statements can be represented as a logical formula composed of atomic statements; the value of validity of each of them is fixed, but in the general case unknown. Then the problem of prediction is reduced to the solution of a system of logical equations and obtaining the values of validity of the atomic statements (or their combinations), and then to the calculation of the validity of new statements according to the available logical formulas. At the same time, it is not always possible to calculate the values of all atomic statements. Therefore, the exact answers can be obtained only for some of the set questions.

*In mathematical statistics* the statistical hypothesis is considered as any statement (assumption) on the general totality verified by sampling. In order to check the hypothesis it is required (in the assumption that it is true) to calculate the value of a certain statistical criterion set as a function of the sampling values. The hypothesis is accepted or rejected depending on whether this criterion exceeds the selected threshold or not. The threshold level is interpreted as the degree of plausibility of the hypothesis. Unlike the problems of classical logics, in this case the statements (hypotheses) are interconnected not by logical operators, but by the sample, in relation to which the decision on their validity is adopted.

At the same time, the task of predicting the results of new measurements is not set at all. By solving the problem of mathematical statistics, we find such a "prefix" of the description of the sample, which would minimize the summary length of the compressed "word" ("prefix" + "suffix"). That is, the second variation problem is solved in accordance with the previously proposed terminology. Thus, we find the most plausible macroscopic description (trajectory in the information space) of the set of results of fundamental measurements. As we have shown previously [28], the principle of minimization of the algorithmic complexity of the description transforms into the ordinary Bayes's formula for the calculation of the criterion of plausibility in case if we are dealing with regularities, which can be set by the function of probabilistic distribution.

The obtained results can be used for the estimate of the results of repeated measurements in a larger sample, however, in this case the changing of the found regularities is not allowed. Therefore, the problems of mathematical statistics can be conventionally called the problems of information statistics. The question on the possibility of prediction of new regularities not contained in the obtained sample is not set in these problems. Even in case when the dynamic characteristics of random processes are analyzed, they are calculated on the basis of the measured interval and approximated without the account of the future changes. At the same time, we are speaking about the physical time scale rather than the information scale, as in our consideration.

By solving the problem of dynamics *in the space of information states,* we obtain one segment of the trajectory of an information object on the basis of the characteristics of its other segment (in the simplest case, these are the initial values of the coordinate and velocity), i.e. we predict the regularities of the results of fundamental measurements obtained in a different moment of information time. In comparison with it, in the problems if classical logics we calculate the results of one fundamental measurements on the basis of the results of the others, in case of presence of strong logical connections between them. At the same time, all a priori information is retained, but the number of its possible interpretations increases. In the problems of mathematical statistics we single out the regular part of the available information and retain it unchanged, leaving a possibility of generating other random parts for it. In both cases we are speaking about the problems of dynamics in a physical sense. Therefore, the model of information space constructed by us allows solving a principally new class of problems of prediction connected with the origin of new information. By calculating the trajectory in the information "future", we in the same way predict the regularities of the set of the results of fundamental measurements after obtaining of additional information. Actually, we predict this new information.

At the same time the variation problem 1 is solved, equivalent to the principle of least action in physics, but not commonly used in classical models.

The following algorithm of its solution can be proposed:

- Selecting the initial language of description of «B» and constructing the scale « ... → $B_{-1}$ → $B_0$ → $B_1$ ... » – a «nest» of information objects, each of which is contained in all subsequent objects and is spaced from the adjacent ones for the same quantity of bits (at the extreme – for 1 bit – one elementary edge). Formally, we can assume as the information object for the language "B" any sequence of its "letters" allowed by the rules of the "grammar".
- By calculating the relative complexity of the set initial information state «$A_0$» (also written in the language «B») and the reference points «$B_i$» of the scale, we can find the minimal upper «$A''_0$» and the maximal lower «$A'_0$» edge of the state «$A_0$».
- By solving *the first variation problem, we* can obtain the information trajectory of the most plausible information "past" (11) and "future" (12) of the state «$A_0$».

Let us note that in case of absence of any "external information forces", due to the principle of inertia the problem of dynamics has a trivial solution – a straight world line in the information space. It corresponds to the conservation of regularities both in case of obtaining additional information and in case of "erasing" of part of the preset information. The principle of "inertia" in the simplest form is used in mathematical statistics.

However, when any information connections occur (in physics it is the kinematic connections or force fields), it appears that the trajectories of motion become curved. The methods of solving the problems of dynamics in physics are mainly intended for calculating such curved trajectories. In the information aspect it means that if any information connections with other objects exist, than their influence on the regularities of description of the considered object, occurring in the process of obtaining additional information, can be taken into account by solving a corresponding problem of information dynamics, i.e. by maximizing the quantity of realizations of these predicted regularities in accordance with the first variation problem and the principle of the "Occam's razor".

### *6.1. Connection of the information time with the physical time*

Let us note that the prediction of the "information future" differs from the prediction of physical future, though it is connected with it. Thus, in the aforesaid example (Fig.3), the information distance $C_3$ is in the information future of the state $D_4$, despite the fact that these two states have been obtained simultaneously. We can only state that if we exclude the possibility of erasing information in the observer's memory, then the sequence of observation of physical events will generate a corresponding scale of information time.

In econophysical applications the role of fundamental measurements is performed by various transactions and the economic state are absolutely ordered of any subsequent state is more preferable compared to the previous one for any of the subjects of such transactions [30]. By selecting a random sequence of absolutely economic states, we select the reference system and the corresponding time scale in the economic space. For a proprietor associated with this reference system the time interval between two points of the scale is measured in the relative increase of his wealth (logarithmic scale). In this case it can also be connected with the scale of physical time – the proprietor will reject all transactions, which are not profitable in his opinion. Therefore, the physical sequence of transactions concluded by him (production cycle, in which the proprietor invests resources with the intention of gaining profit, can also be considered a type of transaction) generates the economic scale of his consecutive economic states.

**Conclusion**

By accepting the "physical" approach to the construction of the fundamental laws of the theory of information, we open a path for constructing it as a dynamic theory, in which, similarly to physics, the notions of space-time, forces, interaction appear. In general, this trend can be called "information physics", similarly to the term "econophysics" used in similar situations in economics. Analysis of the history of development of the physical theory allows us to predict further perspectives of the development of "information physics". Expansion of the set of invariant languages of description in such a way that it would include not only the languages corresponding to the inertial motion, but also the "non-inertial" ones, will likely result in the origin of an information analog of the general theory of relativity. Further expansion of the types of used symmetries is connected with the discreteness of representation of information objects, and, apparently, will require the use of the quantum-mechanical approach.

The question is – whether a finite number of fundamental symmetries which we need to consider for the description of all possible interactions exist, or we are "condemned" to introduce increasingly more complex symmetries for the description of more and more complex systems, remains open in physics as well [7]. In the present publication we have limited ourselves to the analysis of the initial stage of this chain.

Contacts: *smelnyk@yandex.ru*